\documentclass[aps,prb,twocolumn,superscriptaddress,showpacs]{revtex4}
\usepackage{graphicx}
\begin{document}
\title{Homogeneous and inhomogeneous contributions to the luminescence linewidth of point defects in amorphous solids: Quantitative assessment based on time-resolved emission spectroscopy}
\author{Michele D'Amico}\email{damico@fisica.unipa.it}
\affiliation{Dipartimento di Scienze Fisiche ed Astronomiche,
Universit$\grave{a}$ di Palermo, Via Archirafi 36, I-90123 Palermo, Italy}
\affiliation{Istituto di Biofisica, U.O. di Palermo, Consiglio Nazionale delle Ricerche, Palermo, Italy}
\author{Fabrizio Messina}
\affiliation{Dipartimento di Scienze Fisiche ed Astronomiche,
Universit$\grave{a}$ di Palermo, Via Archirafi 36, I-90123 Palermo,
Italy}
\author{Marco Cannas}
\affiliation{Dipartimento di Scienze Fisiche ed Astronomiche,
Universit$\grave{a}$ di Palermo, Via Archirafi 36, I-90123 Palermo,
Italy}
\author{Maurizio Leone} \affiliation{Dipartimento di Scienze Fisiche ed Astronomiche,
Universit$\grave{a}$ di Palermo, Via Archirafi 36, I-90123 Palermo,
Italy}
\affiliation{Istituto di Biofisica, U.O. di Palermo, Consiglio Nazionale delle Ricerche, Palermo, Italy}
\author{Roberto Boscaino}
\affiliation{Dipartimento di Scienze Fisiche ed Astronomiche,
Universit$\grave{a}$ di Palermo, Via Archirafi 36, I-90123 Palermo,
Italy}
\begin{abstract}
The article describes an experimental method that allows to estimate
the inhomogeneous and homogeneous linewidths of the
photoluminescence band of a point defect in an amorphous solid. We
performed low temperature time-resolved luminescence measurements on
two defects chosen as model systems for our analysis: extrinsic
Oxygen Deficient Centers (ODC(II)) in amorphous silica and F$_3^+$
centers in crystalline Lithium Fluoride. Measurements evidence that
only defects embedded in the amorphous matrix feature a dependence
of the radiative decay lifetime on the emission energy and a time
dependence of the first moment of the emission band. A theoretical
model is developed to link these properties to the structural
disorder typical of amorphous solids. Specifically, the observations
on ODC(II) are interpreted by introducing a gaussian statistical
distribution of the zero phonon line energy position. Comparison
with the results obtained on F$_3^+$ crystalline defects strongly
confirms the validity of the model. By analyzing experimental data
within this frame, we obtain separate estimations of the homogenous
and inhomogeneous contributions to the measured total linewidth of
ODC(II), which results to be mostly inhomogeneous.
\end{abstract}
 \maketitle
\section{Introduction}
The physics of color centers embedded in a solid matrix is a
fundamental and interesting scientific field both from the point of
view of basic physics and for their wide technological applications
as modifiers of the macroscopic physical properties of solids, (e.g.
optical transparency, refractive index, electrical resistance, and
so on).\cite{stoneham, nalwa} Several experimental evidences have
led to a general agreement on the fact that the properties of point
defects may be significantly different depending on the crystalline
or amorphous structure of the solid they are embedded
in.\cite{Erice} Indeed, in a crystal each member of an ensemble of
identical defects experiences the same local environment. As a
consequence, every spectroscopical property of the ensemble of
defects, such as the lineshape of the related absorption or
photoluminescence (PL) bands, can be interpreted as a property of
the single center, and is referred to as \emph{homogeneous}.  The
homogeneous absorption linewidth is mainly determined by the
electron-phonon interaction and it is related to other important
physical properties of the defect, such as the Huangh-Rhys factor and
the phonon vibrational frequencies.\cite{nalwa, Erice} On the other
hand, defects in an amorphous solid are believed to feature
site-to-site statistical distributions of the spectroscopic
properties due to the disorder of the surrounding matrix. Hence, the
lineshapes of their optical bands are characterized by an
\emph{inhomogeneous} broadening,\cite{nalwa, Erice, holeburning}
which reflects the degree of disorder of the amorphous solid and
concurs, together with the homogeneous effects, to determine the
overall spectroscopic signature of the color center.

Many experimental approaches have been proposed to estimate the homogeneous and inhomogeneous contributions to the experimental linewidth of an optically active center: exciton resonant
luminescence, resonant second harmonic scattering, femtosecond
photon echo, spectral hole burning and site-selective
spectroscopy.\cite{holeburning,furumiya,woggon,mittlemann,Skujaprb1995,kuroda}
However, the issue is still open since none of these techniques is
applicable to the whole variety of inhomogeneous physical systems of
interest. For instance, in amorphous solids site-selective
spectroscopy can been successfully applied only to defects which
allow the the direct observation of the zero phonon line (ZPL) by
virtue of a weak coupling with the vibrational modes of the
matrix.\cite{Erice,Skujaprb1995}

In this paper we propose a new experimental approach to this
problem, which allows to estimate the inhomogeneous and homogeneous
linewidths based on mapping the variations of the radiative decay
lifetime within an inhomogeneously broadened luminescence emission
band by time-resolved laser-excited luminescence. To this purpose,
in the next section we first describe an adapted version of the
theoretical treatment of the optical properties of a point defect in
a solid, which takes into account the effects of heterogeneity in
amorphous systems. Next, in the experimental section we demonstrate
that the predictions of our model are consistent with the results of
measurements performed on two model point defects, one in a crystal
solid and the other one in a glass. Finally, we use the theoretical
model to estimate the inhomogeneous and homogeneous widths of the
two model defects and to obtain other physical parameters of
interest.

\section{Theoretical description of optical defect
properties} We briefly review the standard theoretical description
of the optical properties of a point defect in a
crystal,\cite{stoneham, nalwa, Erice} in order to adapt it later to
the case of amorphous systems. In addition to the crude
\emph{Born-Oppenheimer} and \emph{Franck-Condon} approximations, we
suppose the defect to be coupled with only one vibrational mode of the solid matrix of frequency $\omega_p$, assumed to be the same for ground and excited electronic states. The frequency $\omega_p$ can be regarded also as the mean frequency of the vibrational modes of the solid or can be thought as the effective phonon frequency coupled with the electronic transition.\cite{nalwa2}

In this frame, the absorption cross section $\Omega(E)$ of a defect
as a function of the excitation energy $E$ at the absolute zero
temperature is given by:\cite{Erice, nalwa2}
\begin{equation}
    \Omega(E)=\beta\sum_k{|M_{0k}|^2} E \cdot \delta[E-(E_0+k \hbar \omega_p)] \label{sigmaE}
\end{equation}
where $\delta$ indicates a Dirac delta function, and the summation
is carried out over the vibronic transitions linking the ground
electronic state with zero phonons towards different vibrational
sub-levels $(k)$ of the electronic excited state, spaced by $\hbar
\omega_p$. The energy value $E_0$ (zero phonon line) is the
absorption transition without emission or absorption of phonons.
$M_{0k}$ is the overlap integral between nuclear wave functions
associated to the ground and excited states, while $\beta$ is given
by: $\beta=\frac{1}{n} \left( \frac{E_{eff}}{E_{ext}} \right)^2
\frac{4 \pi^2}{3 \hbar c}\frac{1}{g_l}|D|^2$, where $D$ is the
matrix element of the electric dipole operator between the ground
and excited electronic states, and $g_l$ is the degeneracy of the
lower electronic state. The effective field correction $\frac{1}{n}
\left(\frac{E_{eff}}{E_{ext}} \right)^2$ accounts for the
polarization effect induced by the external field on the
solid.\cite{stoneham, Erice} We assume here that the refraction
index $n$ is constant in the electromagnetic range investigated. In
the harmonic approximation for the vibrational sublevels relative to
the ground and excited states, the $|M_{0k}|^2$ coefficients are
given by a Poisson distribution:\cite{Erice}
\begin{equation}
    |M_{0k}|^2=e^{-H}\frac{H^k}{k!} \label{Poisson}
\end{equation}
where $H$ is the \emph{Huangh-Rhys factor}, expressing the number of
phonons emitted by the system after absorption of a photon while
relaxing to the ground vibrational substate of the excited
electronic level.\cite{stoneham} Given a population of identical
defects, the envelop of the $\delta$ functions in Eq. (\ref{sigmaE})
describes their characteristic \emph{homogeneous} absorption
lineshape, with (aside from the effect of the factor E) a
$E_{Abs}=E_0+H \hbar \omega_p$ first moment and a
$\sigma_{ho}=\sqrt{H}\hbar \omega_p$ width.

After relaxation towards the bottom of excited electronic state, the
system can relax back to the ground state by spontaneous photon
emission (photo-luminescence). The following relationship of
\emph{mirror symmetry} links the absorption $\Omega(E)$ and
luminescence $L(E)$ band shapes: \cite{Erice}
\begin{equation}
 \frac{L(E)}{E^3}\propto \frac{\Omega(2E_0-E)}{2E_0-E}\label{mirror}
\end{equation}
The energy difference between absorption and emission peaks
(\emph{Stokes shift}) is linked to the Huangh-Rhys factor and results
to be $2S=2H \hbar \omega_p$. Using the \emph{mirror symmetry} Eq.
(\ref{mirror}) and Eq. (\ref{sigmaE}) we obtain:
\begin{equation}
 L(E)\propto \beta \sum_k  |M_{0k}|^2 E^3 \cdot \delta[E-(E_0-k \hbar \omega_p)] \label{PLomog}
\end{equation}
which represents the \emph{homogeneous} emission lineshape, with
(aside from the effect of the factor $E^3$) a $E_{em}=E_0-H \hbar
\omega_p$ first moment and a $\sigma_{ho}$ width. Expression
(\ref{PLomog}) does not take into account the dependence from the
excitation energy within the absorption band. This is based on
experimental results and it will be discussed later.

The PL radiative lifetime $\tau$ is linked to the absorption profile
by the \emph{Forster's equation}: \cite{Erice, forster}
\begin{equation}
 1/\tau=\frac{n^2}{\pi^2c^2\hbar^3} \frac{g_l}{g_u} \int(2E_0-E)^3\frac{\Omega(E)}{E}dE \label{forster}
\end{equation}
where $g_u$ is the degeneracy of the upper electronic state.
Combining Eq. (\ref{forster}) and Eq. (\ref{sigmaE}) we obtain the
decay rate $1/\tau$:
\begin{equation}
 1/\tau=\gamma \sum_k |M_{0k}|^2 (E_0-k \hbar \omega_p)^3 \label{ratetau}
\end{equation}
where $\gamma=\frac{n^2}{\pi^2 c^2 \hbar^3}\frac{g_l}{g_u} \beta $.
The cubic dependence appearing in the above expression is a direct consequence of the relation between Einstein coefficients for absorption and spontaneous emission, which forms the basis of Forster's equation. Eq. (\ref{ratetau}) can be approximated by neglecting the contributions far from $k\sim H$, thus obtaining:
\begin{equation}
1/ \tau = \gamma (E_0-S)^3. \label{tauapprox}
\end{equation}
This expression shows that the decay rate is proportional to
$\gamma$ and approximately to the third power of the first moment of the emission band.

Summing up, the global expression for the luminescence of a
population of identical point defects in a solid matrix as a
function of the spectral position $E$ and time $t$ after an exciting
light pulse (\emph{homogeneous shape}) is:
\begin{equation}\label{rate}
 L(E,t) \propto \gamma \sum_k |M_{0k}|^2 E^3 e^{-t/\tau} \cdot \delta[E-(E_0-k \hbar \omega_p)]
\end{equation}
This expression assumes that non radiative channels from the excited
state are absent. As we see from Eq. (\ref{rate}), the shape and
kinetics of the homogeneous luminescence band are completely
characterized by four parameters: $E_0$ (the ZPL position), $\hbar
\omega_p$ (the phonon energy), $\gamma$ (proportional to $|D|^2$)
and $H$ (the Huangh-Rhys factor). $H$ and $\hbar \omega_p$ can be
expressed in terms of the half Stokes shift $S$ and of the
homogeneous width $\sigma_{ho}$: $\hbar \omega_p$=$\sigma_{ho}^2/S$
and $H=S^2/\sigma_{ho}^2$. In this way, expression (\ref{rate}) can
be alternatively regarded as depending on the four parameters $E_0$,
$S$, $\sigma_{ho}$, $\gamma$, thus being indicated by the
expression: $L(E,t|E_0,S,\sigma_{ho},\gamma)$.

For defects in an amorphous matrix, we can argue the hypothesis of a population of identical defects to fail. Indeed, each point defect interacts with different environments and it is possible that this conformational heterogeneity causes a site-to-site statistical distribution of one or more of the homogeneous properties of single defects. The simplest model we can put forward to take into account the disorder effects is to introduce a gaussian distribution of the ZPL position $E_0$, peaked at $\widehat{E_0}$ and with an inhomogeneous width $\sigma_{in}$; in this scheme, $\gamma$, $S$, and $\sigma_{ho}$ are still considered as undistributed parameters.
Within these hypotheses, the global PL signal $L^*(E,t)$ emitted by the ensemble of non-identical point defects can be now expressed as the convolution of the homogeneous shape $L(E,t)$ with the inhomogeneous distribution of $E_0$:
\begin{eqnarray}
 L^*(E,t|\widehat{E_0},\sigma_{in},S,\sigma_{ho},\gamma)\propto \nonumber \\
 \int L(E,t|E_0,S,\sigma_{ho},\gamma) \cdot e^{-\frac{\left(E_0-\widehat{E_0}\right)^2}{2\sigma_{in}^2}}dE_0 \label{gaussnonomo}
\end{eqnarray}

Eqs. (\ref{rate}) and (\ref{gaussnonomo}) lead us to predict a
difference between the PL signals of defects in crystalline and
amorphous solids. Indeed, when the inhomogeneous broadening
$\sigma_{in}$ is almost zero, as expected for point defects in a
crystalline matrix, Eq. (\ref{rate}) has to be used, and the
radiative lifetime $\tau$ should be independent from the spectral
position at which it is measured within the emission band. In fact,
$\tau$ is expressed by Eq. (\ref{ratetau}), so being a function of
the homogeneous parameters $E_0$, $\gamma$, $S$, and $\sigma_{ho}$,
which are expected to be the same for all defects in the solid. In
contrast, in an amorphous solid a PL band due to an ensemble of
point defects can be thought as arising from the overlap of several
bands with different $E_0$ as described by Eq. (\ref{gaussnonomo}),
and thus featuring different lifetimes. Hence, when $\sigma_{in}$ is
comparable with $\sigma_{ho}$ it should be possible to
experimentally observe a dispersion in $\tau$ by measuring the decay
of the PL signal at different emission energies. Also, the shape of
a band arising from the overlap of sub-bands with different
lifetimes should vary in time, so that the position of its first
moment $M_1(t)$, calculated by the usual expression:
\begin{equation}
M_1(t)=\frac {\int E\ L^*(E,t) dE} {\int
L^*(E,t)dE}\label{firstmoment}
\end{equation} should depend on time. Therefore, both the dispersion of $\tau$
within the emission band and the time dependence of the first
moment can be used in principle as experimental probes of
inhomogeneous effects.

It is worth noting that according to Eq. (\ref{tauapprox}), $\tau$
strongly depends on the first moment of the emission band,
$E_{em}=E_0-S$, and more weakly on $\gamma$. This leads to $E_0$ as
the parameter of choice to be distributed in our model. Moreover, a
gaussian distribution of $E_0$ was experimentally demonstrated for
the non-bridging oxygen hole center point defect in silica, for
which the zero-phonon line can be directly observed by
site-selective spectroscopy at low
temperatures.\cite{Skujaprb1995,vaccaro} On the other side, we
acknowledge that similar predictions can be obtained by introducing
a distribution of the half Stokes shift $S$ with an undistributed
$E_0$. Data reported later on in this paper do not allow to
discriminate between these two possibilities.

Finally, to get further insight into the meaning of Eq.
(\ref{gaussnonomo}) it is useful to consider the extreme case in
which the homogeneous width is so narrow to be negligible with
respect to the inhomogeneous one. In this case, the homogeneous
lineshape $L(E,t)$ can be approximated as $\delta(E-(E_0-S))\cdot
e^{-t/\tau}$, with $\tau$ given by Eq. (\ref{tauapprox}). By
substituting in Eq. (\ref{gaussnonomo}) we get that:
\begin{equation}
 L^*(E,t)\propto e^{-\gamma E^3t}
 \cdot e^{-\frac{\left(E+S-\widehat{E_0}\right)^2}{2\sigma_{in}^2}}
\end{equation}
This expression predicts an exponential decay whose $\tau$ depends
cubically from the experimental observation energy $E$ within the
inhomogeneous band.

In the intermediate situation of non-negligible homogeneous width,
Eq. (\ref{gaussnonomo}) deviates in principle from a single
exponential decay, as it contains contributions with different
values of $\tau$. However, we verified that the typical values of
the parameters which will be used in the following to fit
experimental data ($\widehat{E_0}$, $\sigma_{in}$, $S$,
$\sigma_{ho}$, $\gamma$), correspond to predicted decay curves that
always remain very close to a single exponential for all practical
purposes. From a theoretical point of view, we can define in general
$\tau(E)$ as the time in which $L^*(E,t)$ (at a fixed $E$) decreases
by a 1/e factor from $L^*(E,0)$. With this definition, we can
summarize the above considerations as follows: the $\tau(E)$ curve
(with $E$ varying within the observed emission band) is expected to
vary progressively from a constant value (for a completely
homogeneous system) to a cubic dependence (for a completely
inhomogeneous system) with increasing inhomogeneous/homogenous
ratio. To check the validity of our model we have performed
experimental measurements (described in the following) on
crystalline and amorphous defects.
\section{Materials and Experimental Methods}
We chose F-type-centers in lithium fluoride (LiF) and Oxygen
Deficient Centers of the second type, ODC(II), in amorphous silicon
dioxide (SiO$_2$, or silica) as model point defects on which testing
our approach. Both centers feature broad near-gaussian luminescence
bands in the ultraviolet (UV) range with close decay lifetime values
($\sim$8 ns), and they have both been widely studied in literature
because of their important technological applications. Specifically,
LiF is a material traditionally employed in the production of
high-quality optical elements to be used in the infrared, visible,
and particularly in the ultraviolet spectral regions. F-type-centers
in LiF (electron trapped in anion vacancies) are the subject of
active investigation in the areas of color center lasers, radiation
dosimetry and integrated optics (see Ref. \onlinecite{baldacchini}
and references therein). The study of point defects in silica is a
fundamental technologic problem as well, because their presence
compromises the optical and electrical properties of glasses in
their wide uses as optical components, as insulators in MOS
transistors, and for guiding or processing light signals (optical
fibers and Bragg gratings).\cite{nalwa, Erice} ODC(II) is a peculiar
defect of the amorphous phase of SiO$_2$,\cite{Erice, skujajncs98}
thus being an interesting model system to investigate the
characteristic properties of defects in disordered materials with
respect to crystalline ones. Its microscopic structure consists in
an atom bonded to two oxygen atoms of the matrix
(=X$^{\bullet\bullet}$), where $X$ is an atom belonging to the
isoelectronic series Si, Ge, Sn.\cite{Erice,skujajncs98,skuja1984}
Previous studies have suggested that the spectroscopic properties of
ODC(II) are significantly conditioned by inhomogeneous
effects.\cite{trukhin,leone1, leone2, cannizzophilos}

We report measurements performed on two samples: the first one is a
crystalline Lithium Fluoride sample, hereafter denoted as LiF. Prior
to any measurement this specimen, $5\times5\times1.25$ mm$^3$ sized,
was irradiated at room temperature with electrons of 3 MeV energy,
for a total dose of $1.5\cdot10^6$ rad. The purpose of irradiation
was to induce in the sample the formation of luminescent F-type
centers. The second sample is a fused silica (commercial name:
Infrasil301, provided by Heraeus Quartzglas,\cite{heraeus} and
$5\times5\times1$ mm$^3$ sized), hereafter named I301, manufactured
by fusion and quenching of natural quartz, with typical
concentration of impurities of $\sim$20 ppm in weight.\cite{heraeus}
In particular, as-grown I301 contains a $\sim$1 ppm concentration of
Ge impurities, due to contamination of the quartz from which the
material was produced. Previous studies demonstrated that in the as-grown material most of the Ge impurities are arranged as Ge-ODC(II) defects (=Ge$^{\bullet\bullet}$); moreover comparison with sol-gel silica samples doped with Ge atoms ensures us that the contribution to PL of intrinsic ODC(II) defects in I301 sample is negligible.\cite{skujajncs98,sgjncs03} The optical
activity of Ge-ODC(II) at low temperature ($<$100 K) consists in an
absorption band centered at $\sim5.1$ eV which excites a fast
(lifetime in the ns range) emission band centered at $\sim4.3$ eV,
due to the inverse transition.\cite{skujajncs98,nalwa2,agnelloprb03}

PL measurements were done in a standard back-scattering geometry,
under excitation by a pulsed laser (Vibrant OPOTEK: pulsewidth of 5
ns, repetition rate of 10 Hz, energy density per pulse of
0.30$\pm$0.02 mJ/cm$^2$) tunable in the UV-Visible range. The
luminescence emitted by the sample was dispersed by a spectrograph
(SpectraPro 2300i, PI Acton, 300 mm focal length) equipped with
three different gratings, and detected by an air-cooled intensified
CCD (Charge-Coupled Device PIMAX, PI Acton). The detection system
can be triggered in order to acquire the emitted light only in a
given temporal window defined by its width (t$_W$) and by its delay
t$_D$ from the end of the laser pulse. All measurements reported
here were performed on samples kept at 25 K in high vacuum
($\sim10^{-6}$ mbar) within a He flow cryostat (Optistat CF-V,
OXFORD Inst.). All luminescence signals in I301 were acquired with a
300 grooves/mm grating with a 2 nm bandwidth, while the signals in
LiF were measured with a a 150 grooves/mm grating with a 2.5 nm
bandwidth. All the spectra were corrected for the spectral response
and for the dispersion of the detection system.

\section{Experimental Results}
\begin{figure*}
\includegraphics[width=12 cm]{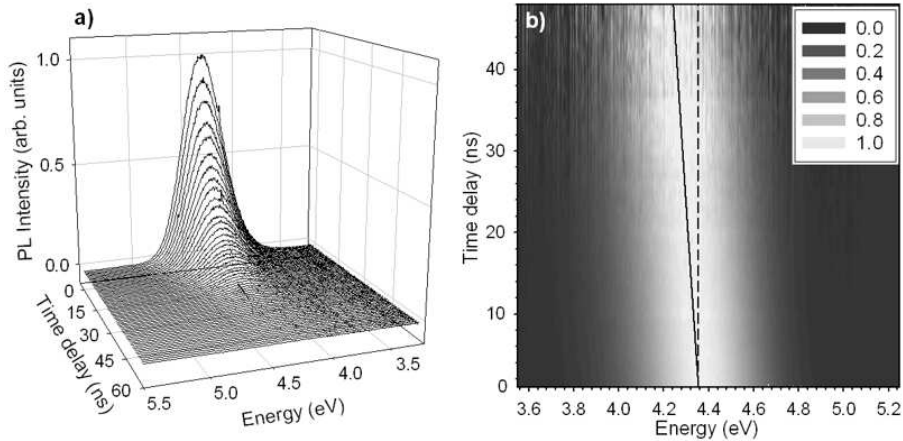} \caption{\label{decad3D}(a) Decay of
the luminescence band of Ge-ODC(II) centers in the I301 sample,
excited at 240 nm at T=25 K. (b) Normalized data of panel (a) in a
contour plot. The continuous line corresponds to the
position of the first moment of the PL band as a function of time. The position of the first moment at $t_D$=0 is reported (dashed line) as a reference.}
\end{figure*}

In Fig. \ref{decad3D}-(a) we show a typical time-resolved
measurement of the PL activity of Ge-ODC(II) in the I301 sample,
performed at 25 K under laser excitation at 240 nm (5.17 eV). The PL
decay was analyzed by performing 60 acquisitions with the same
integration time t$_W$=1 ns but at different delays t$_D$, going
from 0 to 60 ns from the laser pulse. Fig. \ref{decad3D}-(b) shows
the normalized spectra of panel (a) in a contour plot and evidences
that the first moment of the band (continuous line) varies in
time.

\begin{figure}
\includegraphics[width=8 cm]{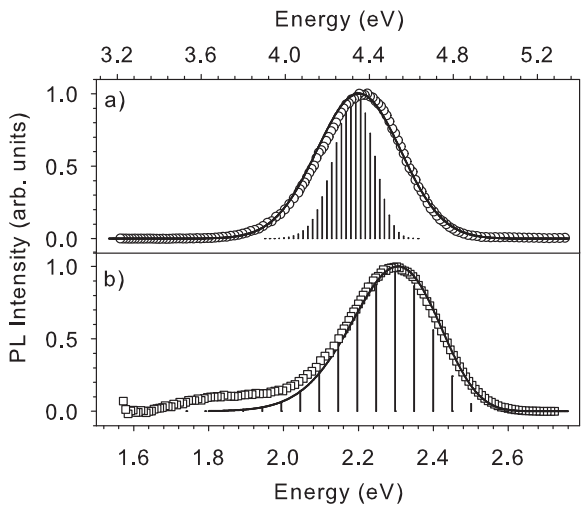} \caption{\label{plblif}Low
temperature (25 K) luminescence of Ge-ODC(II) in the I301 sample
(a), and of F-centers in the LiF sample (b). Both PL bands are
obtained by exciting at the maximum of the respective absorption
bands and acquired for $t_D=0$ and with $t_W$=1 ns. The continuous
line is the result of the fitting procedure by our theoretical
model. The Poissonian homogeneous shape is also shown (see
discussion).}
\end{figure}

In Fig. \ref{plblif}-(a) we report the signal acquired for t$_D$=0,
corresponding to the first spectrum in Fig. \ref{decad3D}-(a). The
PL band of Ge-ODC(II), as acquired immediately after the end of the
laser pulse, is peaked at $\sim$4.4 eV and has a $\sim$0.45 eV width
(Full width at Half Maximum, FWHM) consistent with literature
data.\cite{skujajncs98} Completely analogous time-resolved
measurements were carried out on the PL activity of F-type centers
in the LiF sample.

\begin{figure*}
\includegraphics[width=13 cm]{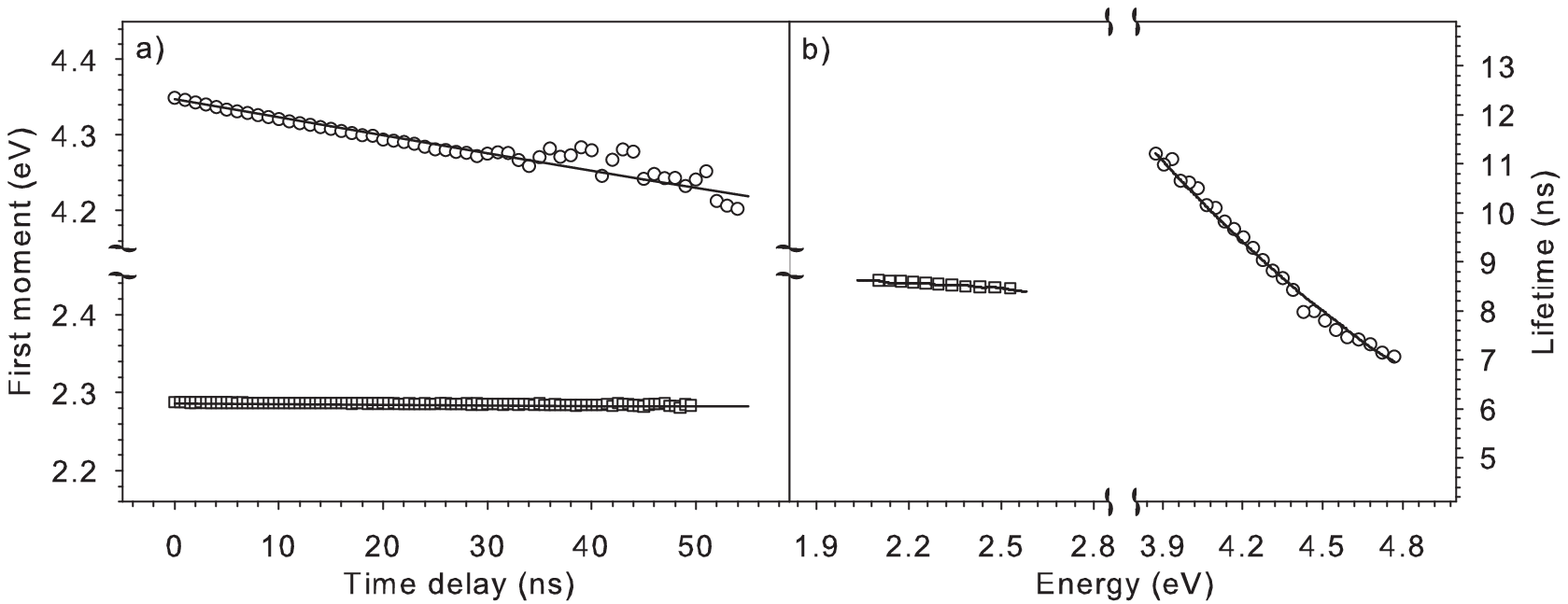} \caption{\label{taumm1}
(a) First moment of the emission band of Ge-ODC(II) (white
circles) and of $F_3^+$ (white squares). (b) Decay lifetime as
estimated by fitting with an exponential function data at different
emission energies within the emission band of the Ge-ODC(II) centers
in the I301 sample (white circles) and of the $F_3^+$ in the LiF
sample (white squares). The continuous lines are the results of the
fitting procedure by our theoretical model (see discussion).}
\end{figure*}

This specimen was excited at 450 nm (2.76 eV) and its luminescence
was collected by varying t$_D$ from 0 to 100 ns with t$_W$=1 ns. We
report in Fig. \ref{plblif}-(b) the luminescence signal detected in
LiF at t$_D$=0. It is apparent that the PL signal of LiF comprises
two contributions peaked at $\sim$2.3 eV and $\sim$1.8 eV. These
signals are known to be associated to two different defects, the
$F_3^+$ and $F_2$ centers respectively, both consisting in
aggregates of F-type centers. \cite{baldacchini,LiF1} In particular,
the main $\sim$2.3 eV band with a $\sim$0.27 eV FWHM is due to
$F_3^+$, consisting in two electrons localized on three adjacent
anion vacancies.\cite{baldacchini} For each activity (Ge-ODC(II) and
$F_3^+$), one can extract the time dependence of the first moment
of the luminescent bands from the time-resolved measurements (e.g.
those in Fig. \ref{decad3D} in the case of Ge-ODC(II)). Data
so-obtained are reported in Fig. \ref{taumm1}-(a). The origin of the
time scale corresponds to t$_D$=0.
We observe that the PL activity in silica shows an approximately
linear decrease of the first moment as a function of time, while
this decrease is not observed in LiF, where the first moment of
the $F_3^+$ centers band has a constant value within experimental
sensitivity. As already discussed in the theoretical section, the
progressive shift of the PL peak position observed for ODC defects
can be alternatively understood as a dependence of the luminescence
lifetime from the spectral position within the emission band. Hence,
in Fig. \ref{taumm1}-(b) we report the values $\tau(E)$ of the PL
lifetime as a function of the emission energy. The lifetimes were
estimated for both PL activities by least-square fitting data from
time-resolved spectra (\ref{decad3D}-(a)) at different emission
energies with an exponential function ($I(t)=
I(0)e^{-t/\tau}$).\footnote{In regard to LiF, the fits were carried
out in the range $\sim$2.10-2.60 eV so as to avoid the region of the
$F_3^+$ emission band possibly affected by the overlap with the
signal due to $F_2$.} At this temperature (25 K), the decays are
purely exponential for both
activities.\cite{agnelloprb03,baldacchini2} Fig. \ref{taumm1}-(b)
shows that the lifetime of Ge-ODC(II) centers in silica strongly
varies within the emission band: $\tau$ goes from $\sim$7.0 ns to
$\sim$10.7 ns. A similar behavior for Ge-ODC(II) was observed also
under excitation by synchrotron radiation.\cite{agnellopss07} On the
contrary, the lifetime of $F_3^+$ centers is almost constant in the
observed range of emission energies. The above results were obtained
exciting at the absorption peak for both PL activities. Although we
performed the same measurements for different excitation energies
within the absorption band, only a very weak dependence from this
parameter was evidenced, consistently with previous
results.\cite{agnelloprb03}

\section{Discussion}
The results in Fig. \ref{taumm1} qualitatively confirm the
predictions of our theoretical analysis, i.e. that the dependence of
the lifetime on the emission energy or, equivalently, the
progressive red-shift of the emission peak with time, are
characteristic features of luminescent defects embedded in a glassy
matrix, as opposed to "crystalline" defects. We stress that the
non-radiative decay channels are almost completely quenched for both
PL signals at the temperature at which the experiments were
performed (25 K).\cite{baldacchini, agnelloprb03} As a consequence,
it is a very good approximation to consider the luminescence decay
to be purely radiative. The main point of the following discussion
is to fit all experimental data by our model and extract the values
of the homogeneous and inhomogeneous widths of the PL emission bands
and other interesting physical parameters.

For both investigated PL activities we have performed numerical
integration of Eq. (\ref{gaussnonomo}) to obtain a set of three
theoretical curves which simultaneously fit i) the shape of the PL
band at t$_D$=0, ii) the time dependence of the first moment
(calculated by Eq. (\ref{firstmoment})) and iii) the dependence of
$\tau$ on emission energy.\footnote{The lifetimes predicted by the
model were estimated by least-square fitting the decay curves (not
reported) predicted by Eq. (\ref{gaussnonomo}) at different emission
energies with a single exponential. It is worth noting that the
simulated data (as real data) feature no appreciable non-exponential
behavior in the timescale of experimental data, at least when the
parameters of the model are close to the best-fit ones.}. To
increase the reliability of the fit procedure, the half Stokes shift
$S$ was fixed to the value obtained experimentally by measuring the
difference between the spectral positions of the absorption and
emission peaks: $S$=0.38 eV and $S$=0.24 eV in silica and LiF
respectively. In this way, the fitting procedure was performed by
varying only four free parameters, $\widehat{E_0}$, $\sigma_{in}$,
$\sigma_{ho}$, $\gamma$. From the experimental point of view, the
vibrational sub-structure of homogeneous luminescence bands cannot
usually be resolved due to the bandwidth of the measuring system and
to further broadening effects due for instance to the coupling with
several low energy modes. To take into account this effect, the
homogeneous lineshape, Eq. (\ref{rate}), was convoluted with a
gaussian distribution of a narrow width $\hbar\omega_p$ before being
inserted into Eq. (\ref{gaussnonomo}).

The continuous lines in Fig. \ref{plblif} and \ref{taumm1} represent
the results of our fitting procedure. It is worth underlining the
goodness of the fit, obtained using only four parameters, and
considering especially that data in Fig. \ref{taumm1} take into
account simultaneously all data acquired in a time-resolved PL
measurement (typically $\sim$600 spectral positions for each of the
$\sim$100 temporal acquisitions of Fig. \ref{decad3D}). Table
\ref{results} summarizes the best parameters obtained via our
fitting procedure for the two investigated PL activities.
\begin{table*}
\caption{Upper section: best fitting parameters obtained by our
theoretical model for the investigated PL activities. Lower section:
Values of $\lambda$, $\sigma_{tot}$, $\hbar \omega_p$, $H$, and $f$,
as calculated from best fitting parameters} \label{results}
\begin{center}
\begin{tabular}{c|c|c|c|c|c}
& $\widehat{E_0}\ [eV]$  & $\sigma_{in}\ [meV]$  & $\sigma_{ho}\ [meV]$  &   $S\ [eV]$  &  $\gamma \ [10^6\ eV^{-3}s^{-1}]$ \\
\hline
I301  & 4.70$\pm$0.05  & 177$\pm$10  &  93$\pm$12  &  0.38$\pm$0.02  & 1.41$\pm$0.09  \\
LiF   & 2.50$\pm$0.02  &  20$\pm$10  &  109$\pm$6  &  0.24$\pm$0.02  & 10.0$\pm$0.6  \\
\hline \hline
& $\lambda (\%)$ &   $\sigma_{tot} \ [meV]$  & $\hbar \omega_p\ [meV]$  & $H$  & $f$\\
\hline
I301 & 78$\pm$5  &  200$\pm$10 &  23$\pm$6 & 17$\pm$5 & 0.073$\pm$0.010 \\
LiF  & 3$\pm$2  &  111$\pm$6 & 51$\pm$7 & 5$\pm$1 & 0.32$\pm$0.04
\end{tabular}\end{center}
\end{table*}
From data in Table \ref{results} we can also calculate the
Huang-Rhys factor $H=S^2/\sigma_{ho}^2$, the vibrational frequency
$\hbar \omega_p =\sigma_{ho}^2/S$, the total width (from
$\sigma_{tot}^2=\sigma_{in}^2+\sigma_{ho}^2$),\footnote{Alternatively,
one can estimate $\sigma_{tot}$ directly from experimental data, so obtaining a consistent value.} and finally the parameter
$\lambda=\sigma_{in}^2/\sigma_{tot}^2$ which estimates the degree of
inhomogeneity. All these quantities are reported in Table
\ref{results} as well. As expected, $\lambda$ is very small for the LiF defects in comparison with the amorphous ones: $\sim3\%$ against $\sim78\%$. These values correspond to $\sigma_{in}$ being about 0.2
times and 2 times $\sigma_{ho}$, in LiF and SiO$_2$ respectively. We note that the inhomogeneous broadening in the crystalline sample is
not exactly zero; beside the approximations in our model, we note
that a real crystal is always distorted by some dislocations,
strains or other imperfections distributed at random into the
matrix. The obtained value of $\lambda$ for Ge-ODC(II) shows that
for a defect embedded in a glassy matrix the inhomogeneous width can
be prominent with respect to the homogeneous one. This conclusion
may be at variance with previous suggestions that $\sigma_{ho}$ and
$\sigma_{in}$ are typically comparable.\cite{skujajncs98}

In Fig. \ref{plblif} we also show the discrete Poissonian
homogeneous lineshape of width $\sigma_{ho}$ and ZPL position
$\widehat{E_0}$, as obtained by our fit procedure for both
investigated activities. As already pointed out, the crystalline PL
band is completely described by the homogeneous shape,\footnote{As
explained above, the homogeneous shape is obtained by a convolution
of the discrete Poissonian with a narrow gaussian curve of width
$\hbar \omega_p$ to take into account further homogeneous broadening
effects and experimental bandwidth.} whereas the silica PL band is
not reproduced without taking into account inhomogeneous effects. It
is also worth noting that the value $\hbar\omega_p$=23$\pm$6 meV
obtained via our fitting procedure is very close to the value of $26
\pm 2$ meV found for the same defect by the analysis of the
temperature dependence of the experimental absorption
linewidth.\cite{cannizzo2003} Moreover, $\hbar\omega_p$ is in good
agreement with experimental and computational works on silica
glasses which predict the presence of vibrational modes of low
frequency.\cite{galeener, umari} These agreements further confirm
the correctness of the present value of $\sigma_{ho}$ and
consequently of our analysis.

\begin{figure}
\centering \includegraphics[width=8 cm]{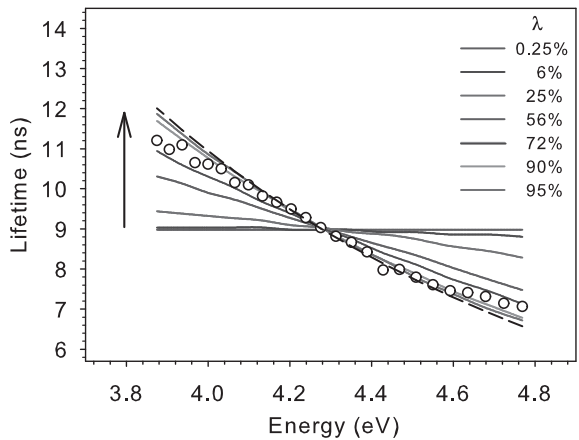}
\caption{\label{fsigmai} Experimental decay lifetime at different
emission energies for Ge-ODC(II) point defects (white circles).
Lifetime as predicted by our model for different values of the
parameter $\lambda$ (continuous lines). Dashed line represents the
extreme case of 1/$E^3$ dependence  (see discussion). The arrow
indicates the direction of increasing $\lambda$.}
\end{figure}

To show the accuracy of our fitting procedure in determining
$\lambda$, in Fig. \ref{fsigmai} we compare the experimental
lifetimes of Ge-ODC(II) in the I301 sample with the predictions of
our model obtained for different $\lambda$ values. The theoretical
$\tau(E)$ curves are obtained by keeping $\sigma_{tot}$ fixed to the
value which best fits the overall experimental shape of the PL band.
This analysis clearly evidences a continuous transition from
constant lifetimes for $\lambda$=0 (that is a completely homogeneous
PL band), to an inverse cubic dependence of $\tau$ from emission
energy for $\lambda$=1 (that is a completely inhomogeneous PL band),
as anticipated in the theoretical section.
Finally, the oscillator strength \emph{f} reported in Table
\ref{results} is calculated using:\cite{Erice}
\begin{equation}
 f=\frac{2m_e}{3\hbar^2e^2} \frac{1}{g_l} E_{Abs}|D|^2\label{oscillator}
\end{equation}
where $m_e$ and $e$ are respectively the mass and the charge of
electron. We have substituted in Eq. (\ref{oscillator}) the value of
$|D|^2$, calculated from the fitting parameter $\gamma$, and we have
used for $E_{Abs}$ the value $\widehat{E_0}+S$. In regard to the
effective field correction, the term $\frac{1}{n} \left(
\frac{E_{eff}}{E_{ext}} \right)^2$ calculated within the Onsager
model,\cite{stoneham, Erice} results to be close to unity both in
SiO$_2$ (n$\sim$1.5) and in LiF (n$\sim$1.4) in the investigated
spectral range. The oscillator strength found here for Ge-ODC(II) in
silica is consistent with the range of values reported in
literature:\cite{skujajncs98} 0.03-0.07. For $F_3^+$ centers in LiF
our result is close to 0.2 reported in ref. \onlinecite{lif2}.

The main assumption of our model that all amorphous effects can be
completely accounted for by a simply gaussian distribution of a
single homogenous parameter (i.e zero phonon line) is strongly
corroborated by the excellent agreement between theoretical curves
and data. On the other side, a distribution of the emission peak
$E_0-S$ is strongly suggested \emph{a priori} by the almost
Einstein-like proportionality of 1/$\tau$ on E$^3$ shown by
experimental data in Fig. \ref{fsigmai}. Finally, it is important to
note that in this scheme $\gamma$ and thus $|D|^2$ are assumed as
undistributed parameters. This means that the oscillator strength
given by Eq. (\ref{oscillator}) can be distributed only as a
consequence of the variations of $E_{Abs}$ associated to different
homogeneous absorption sub-bands.

\section{Conclusions}
We have investigated the inhomogeneous properties of point defects
in a glassy matrix via mapping by time-resolved PL the dependence of
the radiative decay lifetime on emission energy. We propose a
theoretical model, based on an extension of the standard theory of
the optical properties of point defects, incorporating a statistical
distribution of the zero phonon line to account for the effects of
the non-equivalent environments probed by each point defects in an
amorphous matrix as opposed to a crystalline one. This model
enlightens a direct connection between the dispersion of the
radiative decay lifetime within a luminescence band as a function of
emission energy and the inhomogeneous properties of defects in a glassy environment. To confirm our prediction we have experimentally studied the luminescence of Oxygen Deficient Centers in silica and of aggregates of F-centers in a crystalline sample of LiF. The model is able to fit all experimental data and to provide an estimate of
the ratio $\lambda=\sigma_{in}^2/\sigma_{tot}^2$ between the
inhomogeneous and the total width, namely $\sim$78\% for ODCs and $\sim$3\% for F$_3^+$. Finally, our model allowed us to determine the homogeneous parameters of ODC and F$_3^+$ centers: homogeneous width, oscillator strength, Huangh-Rhys factor and the frequency of the vibrational local mode.
\begin{acknowledgements}
We acknowledge financial support received from project "P.O.R.
Regione Sicilia - Misura 3.15 - Sottoazione C". The authors would like to thank R. M. Montereali for having kindly provided the irradiated LiF sample. We also thank G. Lapis and G. Napoli for assistance in cryogenic work. Finally we are grateful to LAMP research group (http://www.fisica.unipa.it/amorphous/) for support and enlightening discussions.
\end{acknowledgements}

\end{document}